\documentclass[journal=jcisd8,manuscript=article]{achemso}

\usepackage{graphicx}
\usepackage{subcaption}
\usepackage{tabularx}
\usepackage{dcolumn}
\usepackage{longtable, booktabs, etoolbox}
\usepackage{bm}
\usepackage{amsmath}
\usepackage{amssymb}
\usepackage{amsthm}
\usepackage{txfonts}
\usepackage{url}
\usepackage[usenames,dvipsnames]{color}
\usepackage{comment}
\usepackage{here}
\usepackage[inline]{enumitem}
\usepackage{lscape}
\usepackage{footnotehyper}
\usepackage{algorithmic}
\usepackage{algorithm}

\newtheorem{prop}{Assumption}

\newcommand{\Aut}{\mathrm{Aut}}
\newcommand{\om}{\Omega}
\newcommand{\ie}{\textit{i.~e.~}}
\newcommand{\argmin}{\mathop{\rm arg~min}\limits}
\newcommand{\shry}{\textsc{SHRY}}
\newcommand{\supercell}{\textsc{Supercell}}
\newcommand{\sod}{\textsc{SOD}}
\newcommand{\ms}{\textsc{Materials Studio}}
\newcommand{\dis}{\textsc{Disorder tool}}
\newcommand{\crystal}{\textsc{CRYSTAL}}
\newcommand{\enumlib}{\textsc{enumlib}}

\newcommand{\pymat}{\textsc{Pymatgen}}
\newcommand{\spglib}{\textsc{spglib}}
\usepackage{booktabs}

\definecolor{Gray}{gray}{0.0}
\definecolor{lightGray}{gray}{0.35}

\author{Genki I. Prayogo}
\email{g.prayogo@icloud.com}
\affiliation[JAIST-mat]
{
  School of Materials Science, JAIST, Asahidai 1-1, Nomi, Ishikawa,
  923-1292, Japan
}

\author{Andrea Tirelli}
\affiliation[SISSA]
{
International School for Advanced Studies (SISSA), Via Bonomea 265, 34136, Trieste, Italy
}

\author{Keishu Utimula}
\affiliation[JAIST-mat]
{
  School of Materials Science, JAIST, Asahidai 1-1, Nomi, Ishikawa,
  923-1292, Japan
}

\author{Kenta Hongo}
\affiliation[JAIST-center]
{
  Research Center for Advanced Computing Infrastructure,
  JAIST, Asahidai 1-1, Nomi, Ishikawa 923-1292, Japan
}
\author{Ryo Maezono}
\affiliation[JAIST-inf]
{
  School of Information Science, JAIST, Asahidai 1-1, Nomi, Ishikawa,
  923-1292, Japan
}

\author{Kousuke Nakano}
\email{kousuke_1123@icloud.com}

\affiliation[SISSA]
{
International School for Advanced Studies (SISSA), Via Bonomea 265, 34136, Trieste, Italy
}

\alsoaffiliation[JAIST-inf]
{
  School of Information Science, JAIST, Asahidai 1-1, Nomi, Ishikawa,
  923-1292, Japan
}

\title{
Application of canonical augmentation to the atomic substitution problem
}

\begin{document}



\begin{abstract}
A common approach for studying a solid solution 
or disordered system within a periodic ab-initio framework
is to create a supercell in which a certain amount of 
target elements is substituted with other ones.
The key to generating supercells is determining how to eliminate symmetry-equivalent structures 
from the large number of substitution patterns. 
Although the total number of substitutions is on the order of trillions, 
only symmetry-inequivalent atomic substitution patterns need to be identified, 
and their number is far smaller than the total.
%
A straightforward solution would be to classify them after
determining all possible patterns, but 
it is redundant and practically unfeasible.
Therefore, to alleviate this drawback, we developed a new formalism based on 
the {\it canonical augmentation}, and successfully applied it to 
the atomic substitution problem.
Our developed \verb|python| software package, which is called 
\shry\ (\underline{S}uite for \underline{H}igh-th\underline{r}oughput 
generation of models with atomic substitutions implemented by p\underline{y}thon),
enables us to pick up only symmetry-inequivalent structures from the vast 
number of candidates very efficiently.
We demonstrate that the computational time required by our algorithm to find $N$ symmetry-inequivalent structures 
scales {\it linearly} with $N$ up to $\sim 10^9$. This is the best scaling for such problems.
\end{abstract}

\section{Introduction}\label{sec.intro}
Atomic substitution on solids 
is the most commonly used strategy in materials science
to tune material properties such as for alloying and cation/anion/vacancy doping, which often demonstrate unprecedented functionalities 
superior to nonsubstituted materials~\cite{2018KAG}.
Within the scientific and technological communities, there is a considerable demand for the ability to predict
the extent to which substitutions affect material 
properties using ab-initio simulations, and to 
evaluate whether they are useful for 
achieving desired properties.
If we consider vacancy-type defects
as a form of substitution (by a vacancy),
this covers another industrially important problem: 
understanding how defects affect materials properties~\cite{2019ICHa, 2019ICHc}. This problem arises during the evaluation of sample qualities, damages, and degradation. 
{\it Ab initio\/} calculations based on density functional theory (DFT) 
are the most promising framework for such an assessment; they utilize  the microscopic structure models of the substitutions. 

\vspace{2mm}
There are many strategies for studying disordered 
crystals or solid solutions within the DFT framework.
If one considers disorder effects at the DFT level
with preserving the periodicity of a target crystal,
virtual crystal approximation (VCA)~\cite{2000BEL}
or coherent potential approximation (CPA)~\cite{1967SOV} 
combined with the Korringa--Kohn--Rostoker (KKR) method~{\cite{1947COR,1954JOH}}
are good choices.
The former achieves it by mixing (all electron or pseudo) potentials
of the constituent elements based on the composition,
whereas the latter utilizes an effective potential of the disordered crystal constructed from the Green's function of the effective medium.
However, because both methods require special implementations, 
not all ab-initio codes can perform them.
The so-called {\it supercell method} is a considerably simpler yet very powerful strategy. 
In this method, substitutions are performed within the supercell of the original (\textit{i.e.}, unsubstituted)
unit cell.
While the supercell approach is simple and applicable for any ab-initio frameworks, an intrinsic problem is that the approach introduces an {\it artificial} periodic-boundary condition that does not exist in real materials.
A direct approach to alleviate this artificial periodicity is increasing the size of the supercell. Although a larger supercell inevitably increases computational costs, recent advances in high-performance computing allow us to handle it even at the phonon analysis level~{\cite{2016NAK,2017NAK}}. Further, a larger supercell affords the finer resolution of the substitution concentration compared to that with an unsubstituted unit cell because only discrete substitutions are allowed in the supercell approach~{\cite{2020UTI}}.
In addition, more sophisticated methods required to alleviate artificial periodicity have been devised.
One such successful method is the quasi-random structures method (SQS)~{\cite{1990WEI, 1990ZUN}}. This method finds the closest periodic supercell to the genuine disordered state based on correlation functions.
If the thermodynamical properties are in focus, the cluster expansion method
with Monte Carlo sampling~{\cite{2009SEK}}
is another very effective technique for studying disordered crystals~{\cite{2021YOS}}.
Several schemes have been developed~{\cite{2009CHR, 2009RUR, 2013MUR, 2014KUM}} to correct spurious interactions between supercell images for studying charged defects in crystals more quantitatively. These methods ensure that the supercell approach is effective and reliable.
In this study, we focus on the supercell approach.

\vspace{2mm}
Within the supercell approach, a solid solution or disordered system is modeled with a supercell in which a certain amount of target elements is substituted with other ones.
The number of possible atomic substitution patterns grows {\it combinatorically} with the substitution concentration and size of the supercell; this readily causes technical issues in the DFT framework.
We consider a magnetic alloy
(Nd$_{0.7}$Ce$_{0.225}$La$_{0.075}$)$_2$Fe$_{14}$B~{\cite{2021UTI}} to illustrate this phenomenon; for this alloy,  a supercell with at least
9 Ce and 3 La substituting
a part of 40 Nd sites is required.
The number of possible
atomic configurations in this case
amounts to
40!/ (28!9!3!) = 1,229,107,765,600~{\cite{2021UTI}}.
However, many of these configurations 
are related with crystallographic symmetry operations; however, only symmetry-inequivalent structures are sufficient for ab-initio simulations.
This implies that 
these {\it redundant} configurations containing many symmetry-equivalent configurations can be further categorized
into those composed of {\it only symmetry-inequivalent configurations}.
The key point here is identifying a set of symmetry-inequivalent configurations
that is vastly smaller than all possible configuration patterns.

\vspace{2mm}
Several software packages have been developed to cope with such vast combinations of atomic substitution in a solid (Table~{\ref{Comparison}}); for example, \supercell~{\cite{2016OKH}}, 
\ms{\textsuperscript{\textregistered}} (\ie, \dis)~{\cite{2018MAT}},
\enumlib~{\cite{2008HAR}} (also combined with \pymat\ (Python Materials Genomics)~{\cite{2013PYM}}),
\crystal~{\cite{2013MUS,2018DOV}}, and
\sod~{\cite{2007GRA}}.
One of the well-known efficient algorithms is the so-called {\it orderly generation}~{\cite{1978RED}} that is employed in \enumlib~{\cite{2008HAR}} and \supercell~{\cite{2016OKH}}.
In this approach, symmetrically redundant structures are eliminated in a backtrack tree search 
by carefully selecting one {\it canonical} representative from every symmetry-equivalent structure set~{\cite{2005KAS}}.
The canonicity of a structure is typically checked using lexicographic orders of permutations.
In this work, we employ the so-called {\it canonical augmentation}~{\cite{2005KAS}}. 
In this approach, structures themselves are not necessarily canonical unlike the orderly generation, 
but the the search tree should be generated {\it in a canonical way}~{\cite{2005KAS}}.
The details are explained later.
The canonical augmentation is similar to the orderly generation 
in the sense that both algorithms exploit canonical labeling map~{\cite{2005KAS}},
but it has several advantages, for instance, one can accelerate the search by using subobject invariants,
as shown later.
Recently, we developed a \verb|python| package, called \shry\ 
(a \underline{S}uite for \underline{H}igh-th\underline{r}oughput
generation of models with atomic substitutions 
implemented in p\underline{y}thon).
\shry\ allows us to pick up only symmetry-inequivalent structures from a set
of redundant structures very efficiently based on the canonical 
augmentation algorithm.
The computational time required by our algorithm to find $N$ symmetry-inequivalent structures scales {\it linearly} with $N$ up to $\sim 10^9$. This is the best scaling for such problems, and superior to existing solutions.
In the following, we describe the package specification,
underlying method, implementation, and several benchmark tests.
%
\begin{table}[htbp]
  \caption{
Comparisons of several programs implementing combinatorial substitutions in crystals
  }
  \begin{center}
    \resizebox{\columnwidth}{!}{%
    \begin{tabular}{cccccccc}
    \hline
    Package & \shry\ & \supercell\ & \ms{\textsuperscript{\textregistered}} & \crystal\ & \sod\ & \enumlib\ \\
    \hline
    Release & 2021 & 2016 & - & 2014 & 2007 & 2013 \\
    Programming language & Python & C++ & Perl & Fortran & Fortran & Fortran + Python \\
    License & MIT & GPL & Commercial & Commercial & GPL & MIT \\
    Input from standard structure files (e.g., CIF) & Yes & Yes & Yes & No & No & Yes \\
    Nondiagonal supercell expansion matrix & Yes & No & Yes & Yes & No & Yes \\
    Coulomb energy sampling & Yes & Yes & Yes & Yes & No & No \\
    Substitutions of several independent Wyckoff positions & Yes & Yes & No & No & No & Yes \\
    Random sampling & Yes & Yes & No & Yes & No & No \\
    \hline
    \end{tabular}%
    }
  \end{center}\label{Comparison}
\end{table}

\section{Algorithms}\label{sec.spec}

\subsection{Summary of technical details}
\shry\ is implemented in Python 3,
and uses the CIF format as the standard
for both reading and writing the reference
and generated sets of substituted structures.
Unlike other packages, \shry\ is purely implemented in \verb|python|,
and therefore, it can be integrated into another \verb|python| program,
in addition to using it as a stand-alone program 
without any complicated compilation procedures.
%
Structure abstractions, manipulations, and symmetry analysis
leverages the feature sets have been reported in \pymat~\cite{2013PYM} and \spglib~{\cite{2018TOG}}.
The key point for the implementation is identifying and avoiding,
among all possible permutations,
multiple instances of symmetrically identical structures.
The implementation of a variant
of the canonical augmentation~\cite{1998MCK} algorithm is unique to \shry\, which allows 
symmetrically unique structures corresponding to
the specified substitute concentration
to be recursively generated from the unique structures.
Furthermore, the entire set of unique structures
for all concentrations can be simultaneously obtained,
which is valuable in the study of increasingly complex alloys.
When the growth in the number of symmetrically unique structures becomes undesirable,
\shry\ also allows the random picking of structures up to a user-specified number.
Polya's enumeration theorem{\cite{1937POL, 2012POL}} provides the expected number
for the irreducible structures to aid in this picking.

\subsection{Theoretical preliminaries}
The redundancy of atomic substitutions is attributed to the existence of symmetries in a crystal. 
Indeed, an atom on a high-symmetric site (\ie, Wyckoff position~{\cite{2016BRO}}) is related to other atoms on the same Wyckoff position via the symmetry operations of the crystal. Thus, one atomic substitution pattern is often equivalent to others from the view point of crystallographic symmetry. \shry\ implements a very efficient technique to avoid the redundancy in this search. 

\vspace{2mm}
Before explaining the algorithm implemented in \shry\, we introduce a few notations and definitions that are used throughout this paper: Given a set $\Omega$, we denote with $\Aut(\Omega)$ the set of bijections $\Omega \rightarrow \Omega$. Let $G$ be a group acting on $\Omega$, \ie, $G$ has a group homomorphism
\[
\phi: G \rightarrow \Aut(\Omega),
\]
called {\it action}. We use the notation $g \cdot X$ for $g \in G$ and $X \in \Omega$ for $\phi(g)(X)$. In this setting, the \textit{orbit of} $X$ is defined as the set 
\[
GX = \{ g \cdot X\ | g \in G \},
\]
and the {\textit{orbit space}} $\om/G$ is the set of all distinct orbits, which is a partition of $\om$. The cardinality of $\om/G$ can be computed using Polya's theorem~{\cite{1937POL, 2012POL}}(sometimes called, Cauchy-Frobenius Lemma, see Ref.~\citenum{2013KER}).
Furthermore, given two elements $X$ and $Y$ in $\om$, the set $\mathrm{Iso}(X, Y)$ is defined as $\mathrm{Iso}(X, Y)=\{g\in G\ |\ g \cdot X=Y\}$ and, when $X=Y$, we shall use the notation $\Aut(X)=\mathrm{Iso}(X, X)$. Note that $\Aut(X)$ is a subgroup of $G$. We will say that two elements are \textit{indistinguishable} if $\mathrm{Iso}(X, Y)\neq \emptyset$, \ie if $X$ and $Y$ belong to the same orbit in $\om/G$. An important role in this paper is played by certain functions on $\om$ whose value is constant over points in the same orbit: given a set $\Gamma$, a function $f: \om \rightarrow \Gamma$ is \textit{invariant} for the action of $G$ if and only if
\[
f(X) = f(g \cdot X), \forall g \in G, \forall X \in \om,
\]
which means $f$ can descend to a well-defined function $\tilde{f}: \om/G \rightarrow \Gamma$, $\tilde{f}(GX) = f(X)$.

\subsection{Atomic pattern substitution search tree}
\begin{figure}[htb]
  \centering
    \includegraphics[width=\linewidth]{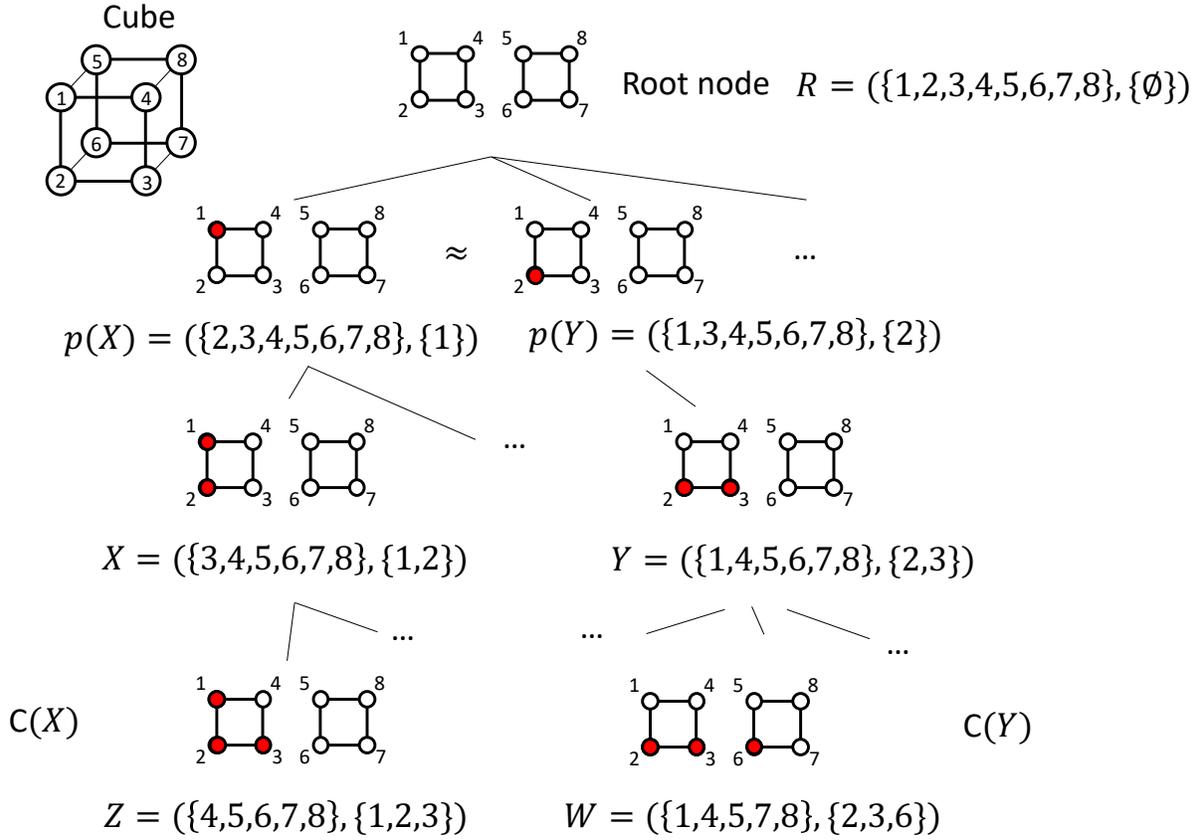}
  \caption{
    Example of the search tree, where three vertices out of eight are colored by red in a cube.
    $p(X)$ denotes the parent node of $X$, whereas the set of children of $X$ are denoted with $C(X)$.
     }\label{fig:tree-example}
\end{figure}

\begin{figure}[htb]
  \centering
    \includegraphics[width=0.7\linewidth]{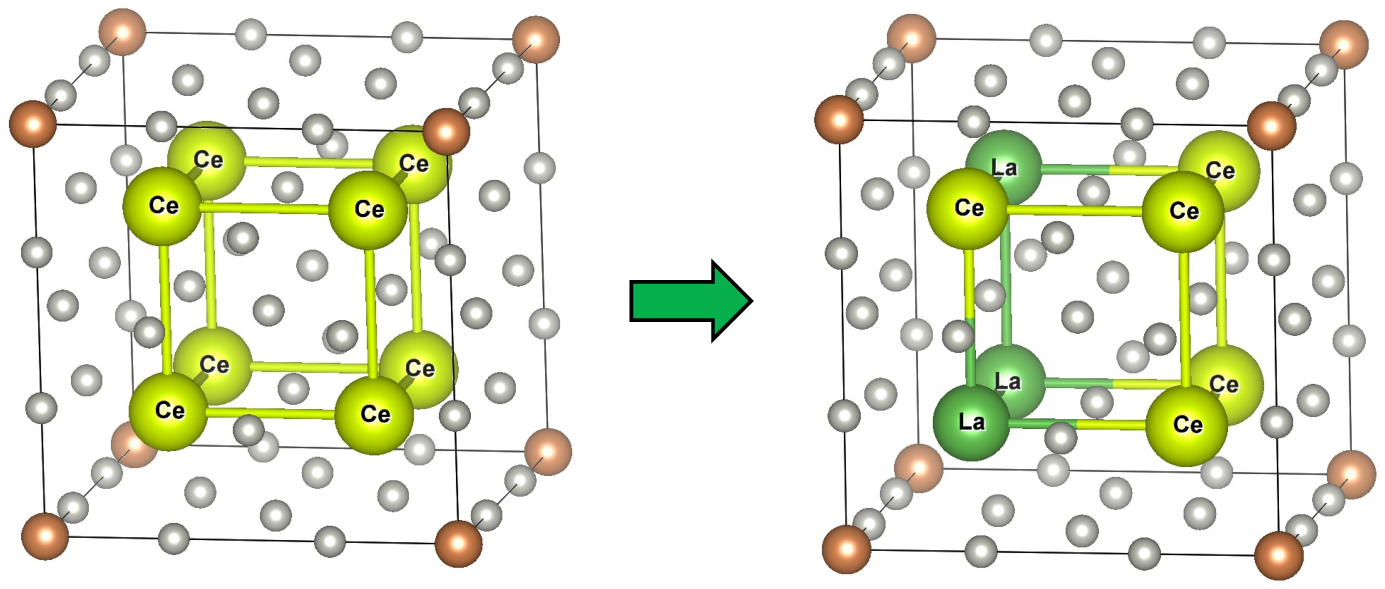}
  \caption{
    Atomic substitutions corresponding to Fig~{\ref{fig:tree-example}}. Ce$_{8}$Pd$_{24}$Sb $\rightarrow$ (Ce$_{5}$,La$_{3}$)Pd$_{24}$Sb, where Ce and La atoms are on the 8$g$ Wyckoff position. 
	The crystal structure~{\cite{1996GOR}} was obtained from the ICSD database~{\cite{1983BER, *2004HEL}} (CollCode: 83378).
	The space group is 221-$Pm\bar{\rm 3}m$; the crystal structures were depicted by VESTA~{\cite{2011MOM}}.
    }\label{fig:ce8pd24sb}
\end{figure}

We consider a very simple case---the problem of coloring the 8 vertices of a cube with two colors, as shown in Fig~{\ref{fig:tree-example}}---to explain the main idea of \shry. We color white vertices with red step-by-step until three of them become red.
This is essentially equivalent to the substitution in which three out of eight atoms on a Wyckoff position are substituted with another element (e.g., see Fig.~{\ref{fig:ce8pd24sb}}. Ce$_{8}$Pd$_{24}$Sb $\rightarrow$ (Ce$_{5}$,La$_{3}$)Pd$_{24}$Sb, where Ce and La atoms are on the 8$g$ Wyckoff position and the space group is 221-$Pm\bar{\rm 3}m$). We discuss atomic substitutions as if there were no translation symmetry, as in the cube example (\ie, not the space group but the point group); however, our considerations can be extended easily to periodic systems.

\vspace{2mm}
The main example of the group action relevant in this work is as follows: let us assign indices $I = (1,2,...8)$ to the vertices of a cube. Let $G$ denote the group composed of symmetry operations of the cube, which is also referred to as the {\it point group} in chemistry (\ie, $O_h$ in the Schoenflies notation). Once we define a one-to-one correspondence between vertices and indices, the action of $g \in G$ on $I$ can be represented as a {\it permutation operation} such as 
$
\begin{pmatrix}
1 & 2 & 3 & 4 & 5 & 6 & 7 & 8\\
8 & 7 & 3 & 4 & 5 & 6 & 2 & 1 \\
\end{pmatrix}
\equiv
(1,8)(2,7),
$
and
$
\begin{pmatrix}
1 & 2 & 3 & 4 & 5 & 6 & 7 & 8\\
7 & 8 & 5 & 6 & 3 & 4 & 1 & 2 \\
\end{pmatrix}
\equiv
(1,7)(2,8)(3,5)(4,6)
$.
Each symmetry of the cube corresponds to one of the permutations. 
For example, the permutation operation (1,8)(2,7) mirrors the cube with respect to the plane containing the vertices \{3,4,5,6\} and (1,7)(2,8)(3,5)(4,6) inverts all the cubes. The cube has 48 automorphisms, \ie, the order of $O_h$ is 48 as there are 48 symmetry operations that do not change the cube.
Therefore, in \shry\, all symmetry operations of a crystal structure associated with the space group are held as a set of permutations of atom indices.

%
%
%

\vspace{2mm}
Let $X$ be an ordered partition of $I$, $X=(V_1, V_2)$,  $V_1\subset I$, $V_2\subset I$, and $V_1 \cup V_2 = I, V_1\cap V_2 = \emptyset$. The sets $V_1$ and $V_2$ are color classes composed of vertex indices. For instance, $X=\{\{3,4,5,6,7,8\}, \{1,2\}\}$ refers to the cube whose first and second vertices are colored by red while the others are white ($X$ in Fig.~{\ref{fig:tree-example}}). We denote with $\Omega$ the finite set composed of all ordered partitions of $I$.
The action of $G$ on $I$ can be easily extended to $\om$ as follows: given $g \in G$ and $X \in \Omega$, we define $g \cdot X$ as 
\[
g \cdot X=(g \cdot V_1, g \cdot V_2),
\]
which is well-defined because $g$ is a permutation and $\{V_1, V_2\}$ is a partition of $I$.

\vspace{2mm}
The set $\om$ of all possible ordered partitions of $I$ (\ie, coloring patterns of $I$) is a key set for our analysis. The elements of $\om$ can be generated recursively using a rooted tree structure $(T_I, R)$ as follows: root $R$ is given by the ordered partition $(I, \emptyset)$ because, for the recursion step, the set of children of $X$, denoted with $C(X)$, is defined as $C(X) = \{ ( V_1 \setminus \{i\}, V_2 \cup \{i\}): i \in V_1 \}$ given a node $X=(V_1, V_2) \in T_I$. There is an urgent need to see that the depth of $T_I$ is exactly $I$. Indeed, we have that the set of nodes on $T_I$ is $\om$; given this identification, we denote a set of non-root nodes of $T_I$ by $\om_{nr}$. Moreover, we can just truncate the tree $T_I$ at depth $|V_2|$ to consider only those ordered partitions where $V_2$ has a fixed number of elements. For a node $X\in T_I$, we denote the parent node of $X$ by $p(X)$. Fig \ref{fig:tree-example} presents the pictorial representation of these definitions. 

\vspace{2mm}
Given this setting, we summarize the main goal to be achieved by \shry\ as follows: efficiently generate all indistinguishable elements of $\om$ by partially traversing the tree $T_I$. That is, we want to extract a representative for each orbit $O\in \om/G$; in this way, we can obtain a fast enumeration of all indistinguishable atomic substitution patterns of a given structure. 

\vspace{2mm}
Although it would certainly lead to a correct algorithm, we cannot hope to achieve our goal through the exhaustive traversal of the search tree $T_I$ because the number of elements of $\om$, \ie\ all possible substitution patterns of a given structure, is of the order of trillions in many cases. Indeed, a straightforward solution to the problem would be to generate all configurations row-by-row, as described by the search tree in Fig.~{\ref{fig:tree-example}}, and then to maintain a record of the complete configurations (\ie~ elements of the set of leaves of the tree $T_I$) encountered, and output a complete configuration only if it is not isomorphic to a complete configuration appearing on record until the number of pooled configurations equals the cardinality of $\om/G$ obtained from Polya's theorem~{\cite{2005KAS}}. Clearly, this solution is correct and works for the simple cube as shown in Fig.~{\ref{fig:tree-example}}; however, for such a solution, it would not be particularly efficient and unfeasible in the actual crystal structure substitution problem, regardless of the implementation, because the total number of patterns often exceeds several billions.
Further, the search tree in Fig.~{\ref{fig:tree-example}} indicates that the same isomorphism class of atomic substitution pattern is encountered multiple times in the traversal; this implies that useless computations are often performed, \ie checking whether the current configuration lies in an orbit that has not been encountered yet. The need of performing all such checks worsens the efficiency of the algorithm, because in many cases, we check if $X\cong Y$ is a computationally demanding operation. 

\vspace{2mm}
\subsection{Isomorph rejection and canonical augmentation}
Efficiency issues outlined above can be addressed using the \textit{isomorph rejection} technique. The term isomorph rejection was introduced in Ref.~\citenum{1960SWI} for techniques that can eliminate redundancy in problems requiring exhaustive search. The main idea of isomorph rejection is explained as follows{~\cite{2005KAS}}: if isomorphic nodes are assumed to have isomorphic children (assumption fulfilled for $T_I$), then, in the traversal of the search tree, we can prune those subtrees whose root belongs to an isomorphism class already encountered; clearly, traversing only the non-pruned subtrees suffices for isomorph-free exhaustive generation. A straightforward implementation of the isomorph rejection involves keeping a global record $\mathcal{R}$ of the objects seen so far during the traversal of the search tree. Whenever an object $X$ is encountered, it is tested for isomorphism against the recorded objects in $\mathcal{R}$. If $X$ is isomorphic to a recorded object, then the subtree rooted at $X$ is pruned as redundant. This approach is presented in Algorithm~{\ref{alg:recorded_object}}.
%
\begin{figure}[!t]
\begin{algorithm}[H]
    \caption{Isomorph rejection via recorded objects{~\cite{2005KAS}}}
    \label{alg:recorded_object}
    \begin{algorithmic}[1]
    \REQUIRE record-traverse($X$)
    \IF{there exists an $Y \in R$ such that $X \cong Y$}
    \RETURN
    \ENDIF
    \STATE $R \leftarrow R \cup \{X\}$
    \PRINT $X$
    \FORALL{$Z \in C(X)$}
    \STATE record-traverse($Z$)
    \ENDFOR
    \RETURN
    \ENSURE
    \end{algorithmic}
\end{algorithm}
\end{figure}
Although the isomorph rejection approach is more efficient than the naive exhaustive generation, the algorithm is not practical for our complex problem because of two reasons: first, a clear difficulty of this implementation is that it requires storing the set $\mathcal{R}$ of the non-isomorphic objects encountered. When the number of nonisomorphic partial objects is large, the available memory (or storage) space can quickly run out. Second, there is still a need to perform the check whether two nodes are isomorphic too many times. 

\vspace{2mm}
These subtleties can be overcome using a technique called \textit{canonical augmentation}, implemented in \shry\, which is a refinement of isomorph rejection. A detailed explanation of this canonical augmentation technique is provided in Chapter 4.3 of Ref.~\citenum{2005KAS}.
In brief, the key of the canonical augmentation algorithm is a relationship between a node $X$ and its parent $p(X)$. To this end, we extend the definition of the isomorphism to the pair of objects: given $X, Y, W, Z \in \om$, $(X,Z)$ is isomorphic to $(Y,W)$, which is denoted as $(X, Z)\cong (Y, W)$ if and only if there exists a $g \in G$ such that $g \cdot X=Y$ and $g \cdot Z=W$. 

\vspace{2mm}
A key feature of the canonical augmentation algorithm is the existence of certain functions that map nodes to other nodes. Consider a function $m: \om_{nr}\rightarrow \om$ such that, for all $X, Y \in \om_{nr}$  $(X, m(X)) \cong (Y, m(Y))$, if and only if $X \cong Y$. The function $m$ is called a {\it canonical parent function} and the pair $(X, m(X))$ is called the {\it canonical augmentation} of $X$. Using such a function $m$, provided that the pair $(\om, m)$ satisfies certain assumptions, one can devise Algorithm \ref{alg:canaug_traverse} so that it returns exactly one element for each orbit of the action of $G$ on $\om$. The assumptions required to ensure that Algorithm \ref{alg:canaug_traverse} returns the desired output are 
\begin{prop}\label{prop:iso_children}
For all nodes $X, Y \in \om$, it holds that, if $X \cong Y$, then for every $Z \in C(X)$ there exists a $W \in C(Y)$ such that $(Z,X) \cong (W,Y)$.
\end{prop}
\begin{prop}
\label{prop:iso_node}
For every nonroot node $X \in \om_{nr}$, there exists a node $Y \in \om_{nr}$ such that $X \cong Y$ and $(Y,m(Y)) \cong (Y,p(Y))$.
\end{prop}
\noindent We say that a node $Z$ satisfying the property of assumption \ref{prop:iso_node}, \ie $(Z, p(Z))\cong (Z, m(Z))$, is {\it generated by canonical augmentation}. The proof that Algorithm \ref{alg:canaug_traverse} is corrected provided Assumptions \ref{prop:iso_children} and \ref{prop:iso_node} are verified can be found in Theorem 4.31 in Ref.~\citenum{2005KAS}. 
%
%
%
%
%
%
\begin{figure}[!t]
\begin{algorithm}[H]\label{algo:canaug}
    \caption{Generation by canonical augmentation{~\cite{2005KAS}}}
    \label{alg:canaug_traverse}
    \begin{algorithmic}[1]
    \REQUIRE canaug-traverse($X$)
    \STATE $A \leftarrow$ Aut($X$)
    \STATE $O \leftarrow \{\emptyset\}$
    \FORALL{$Z$ in $C(X)$}
    \IF{$Z \notin O': \forall O' \in O$}
    \STATE $O \leftarrow O \cup \{C(X) \cap \{aZ: \forall a \in A\} \} $
    \ENDIF
    \ENDFOR
    \FORALL{$O' \in O$}
    \STATE $Z \leftarrow \exists Z \in O'$
    \IF{$(Z, p(Z)) \cong (Z, m(Z))$}
    \STATE canaug-traverse($Z$) 
    \ENDIF
    \ENDFOR
    \RETURN
    \ENSURE
    \end{algorithmic}
\end{algorithm}
\end{figure}
%
\begin{figure*}
  \includegraphics[width=1.0\linewidth]{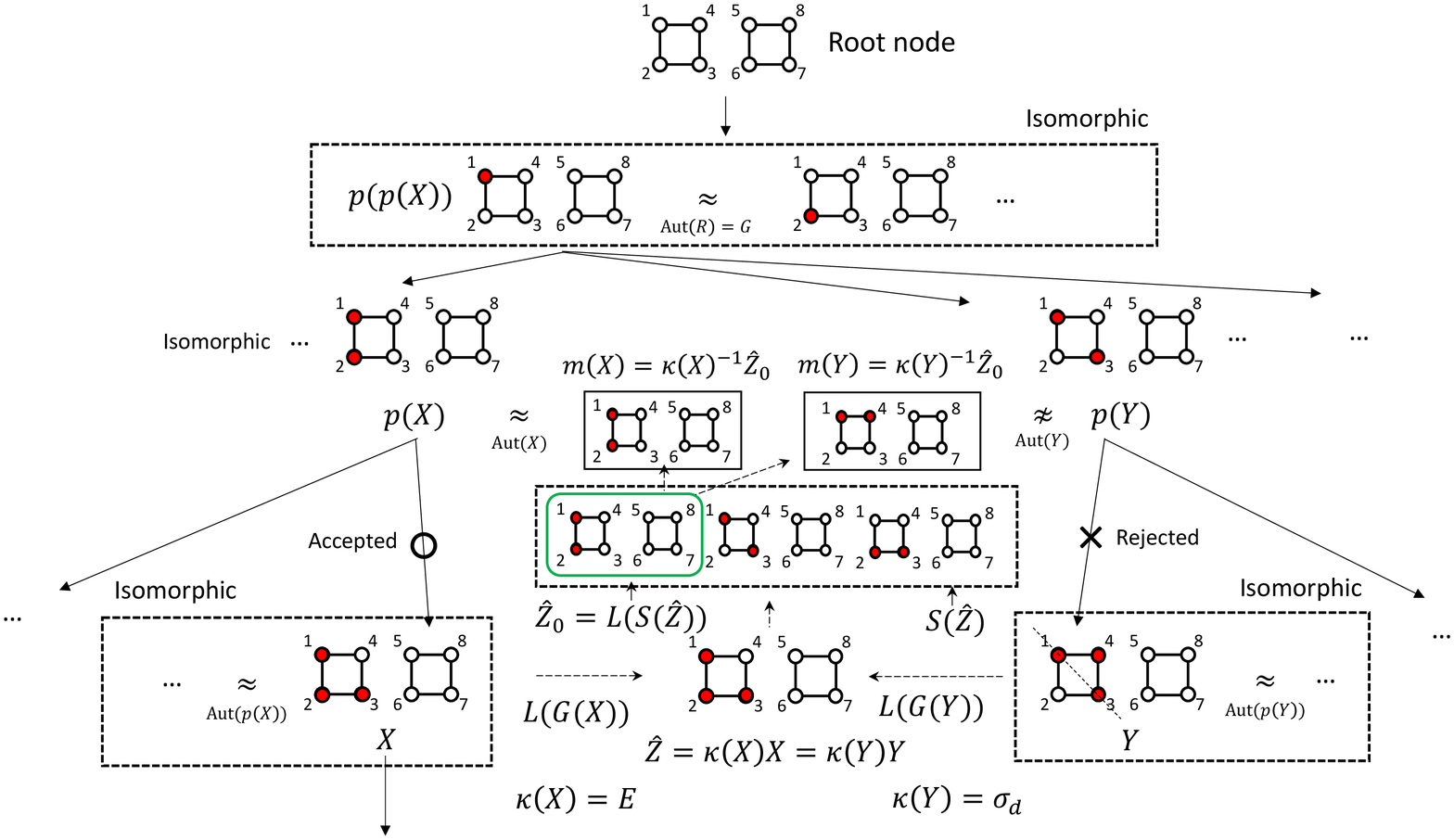}
  \caption{
  A schematic of the generation via canonical augmentation.
  The key to canonical augmentation is the relationship between a node $X$ and its parent $p(X)$. \shry\ checks if $(X,p(X)) \cong (X,m(X))$; otherwise, $X$ is disregarded.
  The selection of a subobject $\hat{Z}_{0}$ depends only on the canonical form $\hat{Z}$ and not on $Z$. Thus, the following important relation holds: $\hat{Z} = \hat{W}$ and $Z_{0} = W_{0}$ for any two isomorphic objects.
  }\label{fig:canaug}
\end{figure*}
%
The key observation on the pair $(\om, m)$ is as follows: assume that $Z$ and $W$ are isomorphic nodes encountered in the tree traversal. Furthermore, assume that both nodes are generated by canonical augmentation. It then follows that 
\[
(Z, p(Z)) \cong (Z, m(Z)) \cong (W, m(W)) \cong (W, p(W)),
\]
which implies $p(Z) \cong p(W)$. Now, if isomorph rejection has already been performed on parent nodes, $Z$ and $W$ are siblings. Moreover, from the above chain of pair isomorphisms, we get that $(Z,X) \cong (W,X)$, where $X=p(Z)=p(W)$ implies that $a \cdot Z = W$ for some $a \in \Aut(X)$. Thus, we conclude that any two isomorphic nodes $Z$ and $W$ generated by canonical augmentation must be related by an automorphism of their common parent node, \ie, $X$. Thus, it follows that one has to perform {\it isomorph rejection only among siblings} when traversing a search tree based on the automorphisms of $X$; this drastically decreases the computational cost compared with the simple isomorphic rejection technique. Furthermore, no objects need to be stored in memory for isomorph rejection. Moreover, the decision whether to accept or reject a node can be made {\it locally} based on a procedure that determines whether the child node $X$ is generated by canonical augmentation. In addition, the search can be efficiently parallelized because disjoint subtrees can be searched independently of each other.

\vspace{2mm}
The key to implementation is defining function $m$ that satisfies the relation $(X, m(X)) \cong (Y, m(Y))$ and Assumptions~{\ref{prop:iso_children}} and {\ref{prop:iso_node}}.
We now explain how to define such a function for the atomic substitution pattern problem, which \shry\ solves.
First, the search tree $T_I$ fulfills Assumption~{\ref{prop:iso_children}}. As such, we are left to finding $m$ such that $m$ is a canonical parent function and Assumption~{\ref{prop:iso_node}} is satisfied.
We define a function $W$ that, given in input a set of non-root nodes, it returns the node that is the smallest with respect to the lexicographic order. We define the function 
\[
L: \mathcal{P}(\om_{nr})\rightarrow \om_{nr},\ L(K) = \argmin K,\]
where argmin works within the lexicographic order, and the notation $\mathcal{P}(-)$ stands for the power set operation. Using $L$, we can define a canonical labeling map for the action of $G$ on $\om$, \ie a function $\kappa: \om \rightarrow G$ such that, for all $g\in G$ and $X \in \om$, the following holds. 
\begin{equation}\label{eq:labelling}
\kappa(gX)gX= \kappa(X)X.
\end{equation}
To define $\kappa$, we proceed as follows: for $X\in\om$, let $g_X\in G$ such that $g_XX = L(GX)$. It is easy to see that, in our case, at least one such $g_X$ always exists, for all $X\in \om$.
Then, we define
\begin{equation}
\kappa(X) = g_X.
\end{equation}
The fact that $\kappa$ is indeed a canonical labeling map is an immediate consequence of the equality \[GX=G(gX),\] that holds for all $g\in G$ and $X\in\om$, from the very definition of the orbit. We need one last definition to define the canonical parent map: for the nonroot node $X$, we define {\it the set of subobjects} of $X$, denoted with $S(X)$, as 
\begin{equation}
S(X) = \{ (V_{1}(X) \cup \{i\}, V_{2}(X) \setminus \{i\}): i \in V_{2}(X)\}.
\end{equation}
The set $S(X)$ can also be considered as the set of all possible parents of $X$ in the search tree $T_I$. Finally, we define the function $m$ as
\begin{equation}\label{eq:m_def}
m(X) = \kappa(X)^{-1}(L(S(\kappa(X)X)))
\end{equation}
The above construction of the map $m$ is an instance of a procedure, outlined in Chapter 4.2.3 of Ref.~\citenum{2005KAS}, to construct a canonical parent function that satisfies Assumption \ref{prop:iso_node} starting from canonical labeling; the procedure reads 
\begin{enumerate}
\renewcommand{\labelenumi}{\alph{enumi}).}
\item Compute the canonical labeling $\kappa(Z) \in G$ and the canonical form $\hat{Z} = \kappa(Z)Z$ for $Z$;
\item Select a sub-object ${\hat Z}_{0} \in S(\hat{Z})$ such that the selection depends only on the canonical form $\hat{Z}$ and not on $Z$ and that the following equation holds: $\hat{Z} = \hat{W}$ and ${\hat Z}_{0} = {\hat W}_{0}$ for any two isomorphic objects, $Z,W \in \Omega_{nr}$;
\item Define $m(Z) \coloneqq \kappa(Z)^{-1} {\hat Z}_{0}$.
\end{enumerate}
Our construction of $m$ follows the above steps; therefore, Assumptions~{\ref{prop:iso_children}} and {\ref{prop:iso_node}} are fulfilled for $m$. Thus, Algorithm~{\ref{alg:canaug_traverse}} allows us to extract representative orbits, \ie, the set of unique atomic substitution patterns, in a very efficient manner.

\subsection{Speeding up canonical augmentation}
\vspace{2mm}
The canonical augmentation algorithm can be further improved using {\it subobject invariants}, as defined in Chapter 3.3.4 Ref.~\citenum{2005KAS}, which allows us to avoid the computationally expensive calculation of the object $m(X)$. 
This is a great advantage of the canonical augmentation algorithm over the orderly generation that needs to check the canonicity of a structure every time~{\cite{2005KAS}}.
Indeed, in Algorithm \ref{alg:canaug_traverse}, we are essentially only interested in testing whether 
\begin{equation}\label{eq:iso_check}
(Z, m(Z)) \cong (Z, p(Z))
\end{equation}
holds for a nonroot node $Z \in \Omega_{nr}$ encountered in the search. It suffices to check whether $m(Z)$ and $p(Z)$ are in the same orbit of the action of ${\rm{Aut}}(Z)$ on the set of subobjects $S(Z)$ (it is necessary to check if the action of $G$ on $\om$ induces an action of $\Aut(Z)$ on $S(Z)$) to check whether the isomorphism Eq.~(\ref{eq:iso_check}) holds. This observation can often be exploited to gain efficiency: indeed, assume that one can define a function $\phi:\om_{nr}\rightarrow \Gamma$, where $\Gamma$ is a poset such that for all $Z \in \om_{nr}$, it induces an invariant map 
\begin{equation}
    \tilde{\phi}: S(Z)/\Aut(Z)\rightarrow \Gamma
\end{equation}
and the minimum of $\tilde{\phi}$ on $ S(Z)/\Aut(Z)$ is unique. Such a minimum is denoted as $\gamma_Z$. 
Within this framework, we replace the definition of $m$ given in Eq.~(\ref{eq:m_def}) with 
\begin{equation}\label{eq:can_aug_mod}
m(X)=\kappa(X)^{-1}(L(\gamma_{\kappa(X)X})).
\end{equation}
With this definition of $m$, it suffices to check whether 
\[
p(X)=\underset{Z\in S(X)}{\argmin}~ \phi(Z).
\]
to verify that $m(X)$ and $p(X)$ are in the same orbit for the action of $\Aut(X)$ on $S(X)$. Indeed, in this case, the property of $\tilde{\phi}$ guarantees that there exists $g\in \Aut(X)$ such that $g \cdot p(X)=m(X)$.
For our case, we consider the function $T:\om_{nr}\rightarrow \mathbb{R}^2$, 
where $T(Z)$ represents a tuple of the sums of the distance matrices, $T(Z) = (D(V); V \in Z$), where
$D(V) = \sum_{i,j}d_{i,j}; i,j \in V$, and $d_{i,j}$ is the Euclidean distance between vertices (atoms) $i$ and $j$ in the cube (in a crystal), and the partial order on $\mathbb{R}^2$ is the lexicographic one.
One can select a different function $D(V)$ as far as it is invariant with respect to symmetry operations such as the determinant, \ie,  $D(V) = \det(d_{i,j}); i,j \in V$.
Only one between $D(V_1)$ and $D(V_2)$ is sufficient to define the invariant in the simple case; however, we use the given definition as it generalizes naturally when substituting atoms in a Wyckoff position with more than two elements, or when targeting more than the Wyckoff positions.
Indeed, function $T$ is invariant with respect to the action of $G$ on $\om$. Therefore, for each $Z\in\om$, $T$ descends to a map $\tilde{T}: S(Z)/\Aut(Z)\rightarrow \mathbb{R}^2$; unfortunately, the minimum of $\tilde{T}$ on $S(Z)/\Aut(Z)$ is not unique; therefore, if we define the canonical augmentation map as per Eq.~(\ref{eq:can_aug_mod}), we cannot straightforwardly follow the above procedure: given two nodes $X$ and $Y$, if $T(X) \neq T(Y)$, $X$ and $Y$ cannot be isomorphic by definition; the vice versa does not hold, \ie two nodes with the same invariant $T$ may or may {\it not} be isomorphic. 
Thus, given a node $Z$, if $p(Z)$ is the unique subobject in $S(Z)$ which realizes the minimum of $T$, then we must have $p(Z) = m(Z)$; this can be decided by invariant computations only, without ever computing $m(Z)$. Similarly, if $p(Z)$ does not have the maximum invariant value in $S(Z)$, $p(Z)$ and $m(Z)$ must occur in different orbits of $Aut(Z)$ on $S(Z)$. In the case where $p(Z)$ is not the unique subobject in $S(Z)$ with the minimum determinant (\ie, $S(Z)$ includes two or more objects which realize the minimum of $T$ in $S(Z)$); then, we must proceed with the usual procedure with canonical labeling and explicitly compute $m(Z)$.
Several benchmark results of the acceleration are shown in Table~{\ref{tab:dmat}}.
The results imply that the use of the distance matrix does not always accelerate the computation.
This is because the current implementation, i.e., the sum of distance matrix, is often too simple 
to distinguish structures belonging to different orbits on a subobject.
A more distinguishable subject invariant could certainly accelerate the exhaustive search.
This is an interesting future work to fully exploit the advantage of the canonical augmentation algorithm.
%
\begin{figure}[htb]
  \centering
    \includegraphics[width=0.8\linewidth]{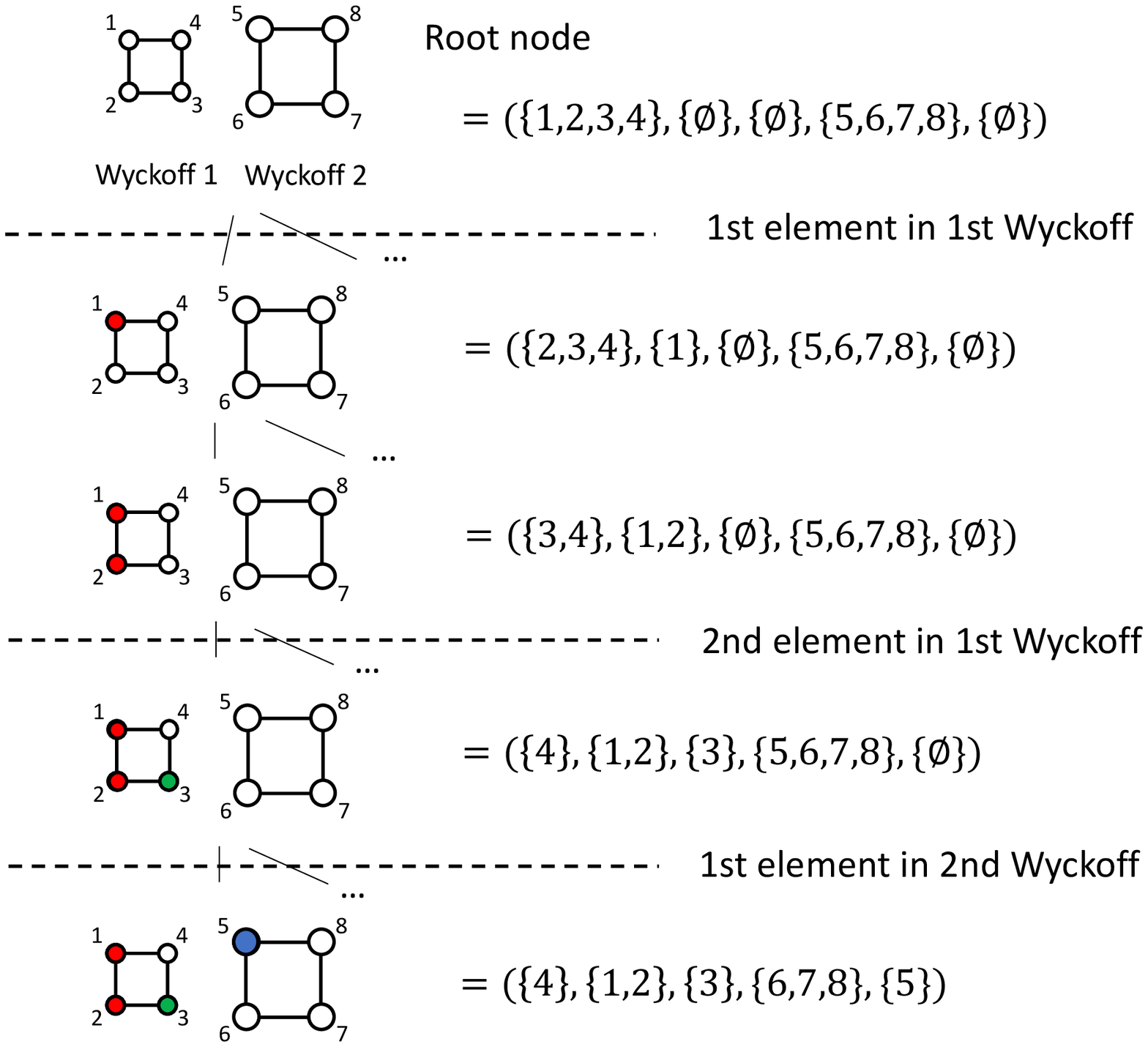}
  \caption{
    An extension of the search tree to replace a single site with three or more elements and/or to replace two or more sites. When the atomic substitution ends for the $j$-th element in the $i$-th Wyckoff position, \shry\ proceeds with the substitution for the $(j+1)$-th element in the $i$-th Wyckoff position. Similarly, when the atomic substitution ends for the $i$-th Wyckoff position, \shry\ proceeds with the substitution for the $(i+1)$-th Wyckoff position.
    }\label{fig:tree-multi}
\end{figure}
%

\vspace{2mm}
\subsection{Extensions}
The algorithm can be easily extended to replace a single site with three or more elements or to replace two or more sites.
We assign sequential indices for the atomic site $I=\{1,2,....N_{\rm atom}\}$, where $N_{\rm atom}$ is the number of atoms in the simulation cell. Then, we define
\begin{equation}
X=(V_{1,1}, \cdots, V_{1,n_c(1)}, \cdots, V_{i,1}, \cdots, V_{i,n_c(i)}, \cdots, V_{n_w,1}, \cdots, V_{1,n_c(n_w)}),
\end{equation}
where $n_c(i)$ and $n_w$ represent the number of different elements in the $i$-th Wyckoff position, and the number of Wyckoff positions in a crystal, respectively. The same element in different Wyckoff sites are treated as different type of elements. In the multielement and multi-Wyckoff frameworks, an analogue of tree $T_I$ can be generated according to the rule: given a node $X$, the set of its children is $C(X)$, which is defined as
\begin{equation}
\begin{split}
C(X) = \{ (
& V_{1,1}, \cdots, V_{1,n_c(1)}, \cdots, \\
& V_{i,1} \setminus \{l\}, \cdots, V_{i, j} \cup \{l\}, ... \\
& V_{n_w,1}, \cdots, V_{n_w,n_c(n_w)}
; l \in V_{i,1}
) \}
\end{split}
\end{equation}
for the $j$-th ($j>1$) element in the $i$-th wyckoff position. When the atomic substitution ends for the $j$-th element in the $i$-th Wycoff position, \shry\ proceeds with the substitution for the $(j+1)$-th element in the $i$-th Wyckoff position.
{\footnote{One can employ a different sequence of the substitution: e.g. when substituting A with B and C (A$_8$ $\rightarrow$ A$_5$B$_2$C$_1$), substituting A with B' (=B+C) (i.e., A$_8$ $\rightarrow$ A$_5$B'$_3$) first, then replacing B' with C (i.e., A$_5$B'$_3$ $\rightarrow$ A$_5$B$_2$C$_1$).}}
Similarly, when the atomic substitution ends for the $i$-th Wyckoff position, \shry\ proceeds with the substitution for the $(i+1)$-th Wyckoff position. The schematic is shown in Fig.{~\ref{fig:tree-multi}}.
Accordingly, given a node $X$, the set $S(X)$ of subobjects of $X$ is defined with the formula 
\begin{equation}
\begin{split}
S(X) = \{ (
& V_{1,1}, \cdots, V_{1,n_c(1)}, \cdots, \\
& V_{i,1} \cup \{l\}, \cdots, V_{i, j} \setminus \{l\}, ... \\
& V_{n_w,1}, \cdots, V_{n_w,n_c(n_w)}
; l \in V_{i,j}
) \},
\end{split}
\end{equation}

for the $j$-th ($>1$) element in the $i$-th Wyckoff position.
Functions $L$, $K$, and $T$ are extended in a natural manner. Thus, theorems and methods discussed for the single-element, single-Wyckoff case also hold for the multielement and multiWyckoff cases.

\vspace{2mm}
Another possible extension of the canonical augmentation algorithm is to the case of periodic systems. This can be achieved by using the space group $\tilde{G}$ instead of the point group ${G}$. A user needs to specify the size of the supercell so that the number of substituted atoms becomes an integer for the desired solid solubility. The space group $\tilde{G}$ is composed of an infinite number of elements of translation symmetry. Therefore, we treat all translation symmetries larger than the lattice vectors of the specified supercell size as the identity transformation.

\begin{center}
\begin{table*}[htbp]
  \caption{
    Comparison of pattern generation times
    with and without the distance-matrix acceleration technique.
    Here, Code refers to the unique ID of the corresponding reference structure in Crystallography Open Database (COD)~{\cite{2009GRA}}.
    The CPU times were measured on 
    Intel{\textsuperscript{\textregistered}} Xeon{\textsuperscript{\textregistered}} G-6242 (2.8~GHz) processor
    by the {\textsc{time}} command, 
    where the time-consuming I/O operations were disabled.
  }\label{tab:dmat}
    \resizebox{\columnwidth}{!}{%
    \begin{tabular}{cccc|cc|cc|c}
    \hline
    Compound & Code & Supercell & Composition & Patterns & Symmetries & With~(s) & Without~(s) & speed up \\
    \hline
    (Pb,Sn)Te$_{2}$ & 9011358 & 2 $\times$ 2 $\times$ 2 &  Pb$_{5}$Sn$_{11}$Te$_{32}$    & 45,497 & 1536 & 13.74 & 38.92 & $\times$ 2.83 \\
    Bi$_{2}$CuO$_{4}$ & 1008474 & 2 $\times$ 1 $\times$ 1 &  Bi$_{16}$Cu$_{8}$O$_{24}$S$_{7}$Se$_{1}$ & 5,259,150 & 16 & 133.41 & 308.81 & $\times$ 2.31 \\
    SmFe$_{12}$     & 1525023 & 1 $\times$ 1 $\times$ 1 &  Sm$_{2}$Fe$_{15}$Ti$_{6}$Mo$_{3}$ & 148,176 & 256 & 113.83 & 114.85 & $\times$ 1.01 \\
	\hline
    \end{tabular}
    }
    \end{table*}
  \end{center}

\subsection{Coulomb energy calculations}
Coulomb energy is useful for predicting
the stability of the proposed structure
before performing more
expensive \textit{ab initio} calculations.
Since \shry\ can provide 
Pymatgen's \texttt{Structure} in runtime,
a user can filter out less stable structures
before saving them to the disk
using the corresponding module from Pymatgen
or any other tool of choice.
We provide an example
on how to do this with
the distribution.

\section{Comparison with existing software packages}
\vspace{2mm}
Since the atomic substitution problem is very general and important, several software 
packages have already been developed. Table~\ref{Comparison} presents a comparison of \shry\ 
with the existing software packages.
%
To our best of knowledge, the site-occupancy disorder (\sod) package is the pioneering work in this field {\cite{2007GRA}}. 
The software package has been applied for many
disordered systems so far since 2007.
Another open-source package with similar functionalities is the \enumlib\ package~{\cite{2008HAR}}.
The code enables us to generate derivatives superstructures of a parent lattice in a considerably efficient manner, and it has been integrated into \pymat~{\cite{2013PYM}}. 
The atomic substitution problem is solved by several commercial software packages. 
The \dis\ tool implemented in \ms{\textsuperscript{\textregistered}}~{\cite{2018MAT}}
is a very user-friendly GUI-based module that is used to generate solid-solutions.
Similar functionalities are implemented in a well-known DFT code, \crystal~ program~{\cite{2018DOV, 2020DOV}}, to treat disorder systems within the periodic ab-initio framework.
\supercell~{\cite{2016OKH}} is the most recent and sophisticated software package in the field. The package implemented a considerably efficient algorithm for the atomic substitution problem called ``orderly generation,''~{\cite{1978RED, 2005KAS}}, which is similar to our canonical augmentation implementation in the sense that both algorithms exploit canonical map labeling~{\cite{2005KAS}}.
Within the existing solutions, Okhotonikov et al.~{\cite{2016OKH}} reported that 
only the \supercell\ package could deal with cases with a large number of 
permutations because the other programs would crush on such computationally demanding test cases. 
Therefore, for the benchmark test, the relevant comparison is between \shry\ ver.~1.0.0 and \supercell\ ver.~1.2.

\section{Benchmark test}
A$_{x}$Pb$_{1-x}$Te is a typical benchmark system used for testing the performance of an atomic substitution program. Table~\ref{table_PbTe} summarizes the results for this benchmark test. Since \shry\ is implemented in \verb|python| (\ie an interpreted language), 
it is intrinsically slower than the other ones implemented in compiled programming languages such as C++. Nevertheless, \shry's performance for the largest problem ($N$ $\sim$ 400,000) 
is as fast as the \supercell\ implemented in C++.

\begin{figure}[htb]
  \centering
    \includegraphics[width=1.0\linewidth]{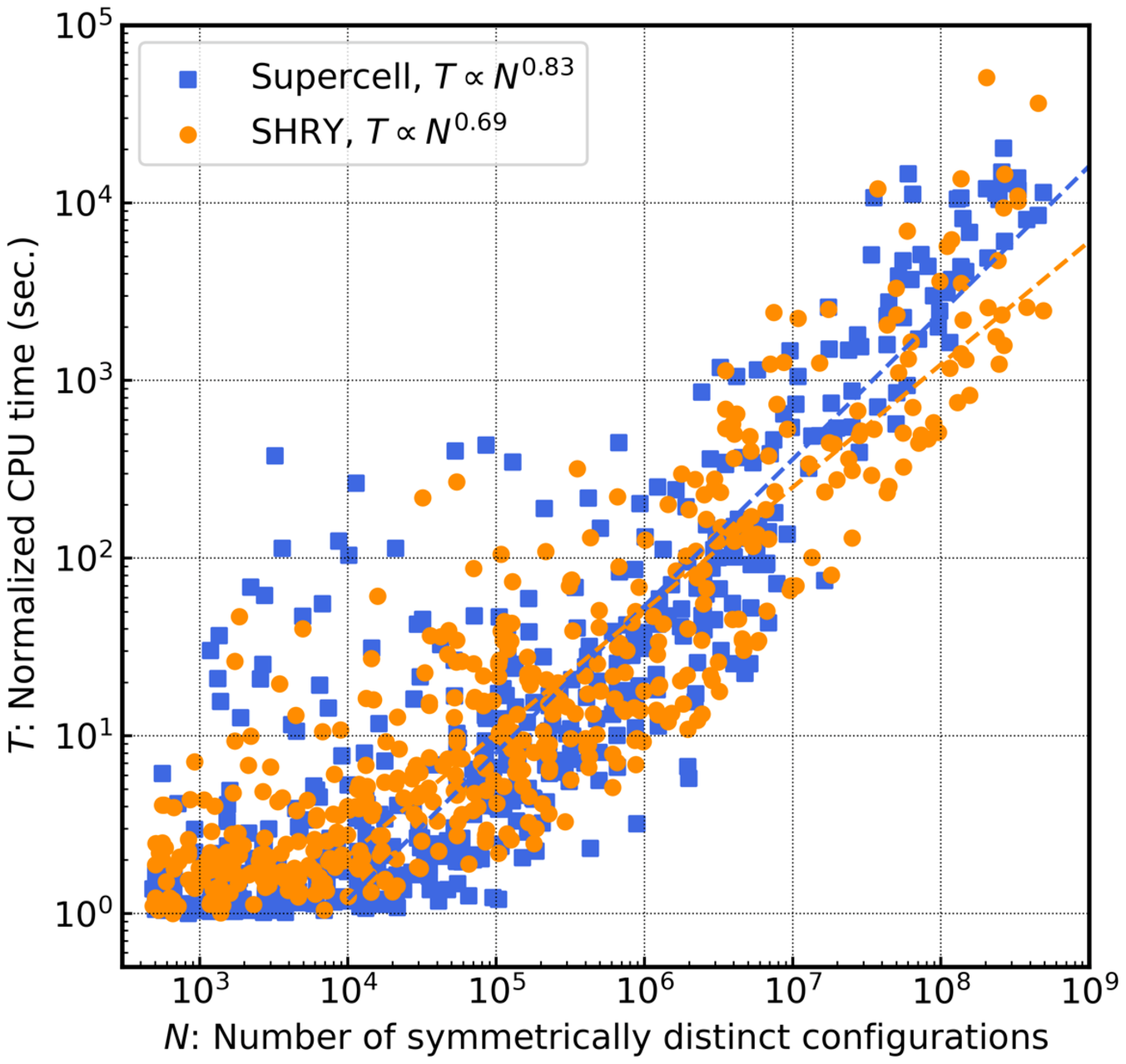}
  \caption{
    Comparing the normalized CPU times of \shry\ and \supercell\ on a benchmark dataset. On the horizontal axis, we place the number of symmetrically distinct structures whereas the vertical axis indicates the normalized CPU time. The linear fits, shown by the broken lines, were performed using only the $N \ge 10^4$ points.
    }\label{fig:shry-supercell-scaling}
\end{figure}

\vspace{2mm}
We benchmarked \shry\ and \supercell\ for a variety of systems.
For the benchmark data set, we randomly chose several \verb|cif| files from Crystallography Open Database (COD)~{\cite{2009GRA}} for each space group, as listed in Table~\ref{table_validation} (Supporting Information), and we generated \verb|cif| files containing up to three atomic substitution sites.
The supercell size and atomic substitution sites were randomly and repeatedly selected for each compound 
until the total number of symmetry-inequivalent structures becomes less than $10^9$.
The number of compounds used for the benchmark test is 500.
Table~\ref{table_validation} lists the compounds, unique IDs in the database, space groups, compositions used for the benchmark test. Further, we report the sizes of supercells, substituted Wyckoff position(s), total numbers of substituted structures, and numbers of symmetry-inequivalent structures.
The computational times required to find all symmetry-inequivalent structures are summarized in the rightmost columns of Table~\ref{table_validation}.
We confirmed that the total amount of substitution patterns including symmetry-equivalent structures and number of symmetry-inequivalent structures are consistent between \shry\ and \supercell\ for all compounds listed in Table~\ref{table_validation} with the given substitution patterns.
Since \supercell\ is written in a compiled language (C++), while \shry\ is implemented in \verb|python|, we performed a normalization on the total CPU times to {\it fairly} compare the performances: namely, given a set of validation structures, $S$, and by denoting the CPU times taken on configuration $i \in S$ by \shry\ and \supercell\ as $t_{shry}(i)$ and $t_{supercell}(i)$, respectively, we define the normalized times $\tilde{t}_{shry}(j)$ and $\tilde{t}_{supercell}(j)$ as 
\[
\tilde{t}_{*}(j) = \frac{t_{*}(j)}{\min_{i\in S}t_{*}(i)}, 
\]
for $*=shry,\ supercell$.
In Fig. \ref{fig:shry-supercell-scaling}, we show the comparison, which indicates that, in this normalized setting, 
the computational time is almost constant up to $N \sim 10^4$ in the small $N$ region both in \shry\ and \supercell; here, $N$ refers to the number of symmetrically distinct configurations. This is because preconditioning parts such as finding symmetry operations of an unsubstituted structure are the rate-determining steps of the algorithms (e.g., the actual CPU times are less than 1 s and 3 s for $N \le 10^4$ in \supercell\ and \shry, respectively). Instead, the computational time in the large $N$ region increases as $N$ increases; however, it scales {\it linearly} up to $N \sim 10^9$ both in \shry\ and \supercell, \ie, they achieve the best scaling for such problems. The linear fits with the $N \ge 10^4$ points show that the normalized computational times scale with $N^{0.83}$ and $N^{0.69}$ for \supercell\ and \shry, respectively.
\begin{table*}[t]
  \caption{
The total number of possible atom combinations for different supercell sizes of the A$_{0.5}$Pb$_{0.5}$Te system. 
    The CPU times of \shry\ and \supercell\ were measured on 
    Intel{\textsuperscript{\textregistered}} Xeon{\textsuperscript{\textregistered}} G-6242 (2.8~GHz) processor 
    by the {\textsc{time}} command, where the time-consuming I/O operations were disabled. The distributed binary
    \supercell\ program was used for the test. 
  }
  \begin{center}
    \resizebox{\columnwidth}{!}{%
    \begin{tabular}{cccccc}
    \hline
    Total number of atoms in the simulation cell & 8 & 16 & 24 & 32 & 64 \\
    \hline
    Supercell size $a \times b \times c$ & $1 \times 1 \times 1$ & $1 \times 1 \times 2$ & $1 \times 1 \times 3$ & $1 \times 2 \times 2$ & $2 \times 2 \times 2$ \\
    Symmetry operations & 192 & 128 & 192 & 256 & 1536 \\
    Total combinations (symmetry-inequivalent structures) & 6(1) & 70(8) & 924 (34) & 12,870 (153) & 601,080,390 (404,582) \\
    Performance of \shry~(s) & 1.31 & 1.32 & 1.45 & 1.64 & 252.69 \\
    Performance of \supercell~(s) & 0.39 & 0.40 & 0.46 & 0.57 & 408.84 \\
    \hline
    \end{tabular}
    }
  \end{center}
  \label{table_PbTe}
\end{table*}

\section{Conclusion}
\label{sec.conc}
We applied isomorph-free exhaustive generation 
based on canonical augmentation to the atomic substitution problem.
The canonical augmentation allowed us to pick up
only symmetry-inequivalent structures from the vast number of candidates
very efficiently. This is very useful when employing the supercell approach 
to study a solid solution or a disordered system in an ab-initio calculation.
The proposed algorithm \shry\ (\underline{S}uite for \underline{H}igh-th\underline{r}oughput generation of models with atomic substitutions implemented by p\underline{y}thon) was implemented in a python software package.
Although we showed several examples where some elements are substituted 
by other elements in this paper, the application of \shry\ is not limited to them.
For instance, \shry\ can also be applied for studying vacancies, magnetic structures 
(e.g., configurations of up and down spins), and charge disproportionation in crystals 
within the supercell approach as far as symmetry-inequivalent patterns are concerned.
A great advantage of the canonical augmentation algorithm over the other isomorphic rejection techniques
is the use of the subobject invariant that enables us to avoid computing the time-consuming canonical labeling map.
The present implementation, i.e., the sum of the distance matrix, is sometimes not capable of distinguishing objects 
belonging to different orbits on a subobject. Finding a more distinguishable subject invariant is an interesting future work to fully exploit the advantage of the canonical augmentation algorithm, realizing a more efficient exhaustive search.

\section{Data and Software Availability}
The python software package, \shry, is distributed on our {\textsc{GitHub}} repository [\url{https://doi.org/10.5281/zenodo.5652360}] under the MIT license.
All the CIF files used for the benchmark test are also available from our {\textsc{GitHub}} repository [\url{https://doi.org/10.5281/zenodo.5652360}] and those of unsubstituted structures are available from Crystallography Open Database (COD)~{\cite{2009GRA}} [\url{https://www.crystallography.net/cod/}]. The unique IDs of the compounds in the database, the compositions and the substituted Wyckoff position(s) are shown in Table~\ref{table_validation}.
The measured CPU times plotted in Fig.~{\ref{fig:shry-supercell-scaling}} are summarized in the rightmost columns of Table~\ref{table_validation}.

\section{Acknowledgments}
The authors acknowledge the facilities of 
Research Center for Advanced Computing 
Infrastructure at JAIST. 
G.P. gratefully acknowledges the financial support from the JST SPRING (Grant Number JPMJSP2102).
A.T. acknowledges financial support from the MIUR Progetti di Ricerca di Rilevante Interesse Nazionale (PRIN) Bando 2017 (Grant Number 2017BZPKSZ).
K.H. is grateful for financial support from 
the HPCI System Research Project (Project ID: hp210019, hp210131, and jh210045) and 
MEXT-KAKENHI (JP16H06439, JP17K17762, JP19K05029, JP19H05169, and JP21K03400)
and the Air Force Office of Scientific Research
(Award Numbers: FA2386-20-1-4036).
R.M. is grateful for financial supports from 
MEXT-KAKENHI (JP16KK0097, JP19H04692, and JP21K03400), 
FLAGSHIP2020 (project nos. hp190169 and hp190167 at K-computer), 
the Air Force Office of Scientific Research 
(AFOSR-AOARD/FA2386-17-1-4049;FA2386-19-1-4015), 
and JSPS Bilateral Joint Projects (with India DST). 
K.N. acknowledges a support from the JSPS Overseas Research Fellowships, that from Grant-in-Aid for Early-Career Scientists (Grant Number JP21K17752), and that from Grant-in-Aid for Scientific Research(C) (Grant Number JP21K03400).


\bibliography{references}



\setcounter{table}{0}
\setcounter{equation}{0}
\setcounter{figure}{0}
\renewcommand{\thetable}{S-\Roman{table}}
\renewcommand{\thefigure}{S-\arabic{figure}}
\renewcommand{\theequation}{S-\arabic{equation}}


\begingroup
\fontsize{6.5pt}{0.2cm}\selectfont
\begin{center}
  \begin{landscape}
\begin{longtable}{c|c|c|c|c|c|c|c|c|c}
    \caption{\label{table_validation} 
    The compounds used for the benchmark test. The original cif files were downloaded from Crystallography Open Database (COD)~{\cite{2009GRA}}.}\\
    \hline
    Reduced formula & Code{\footnotemark[1]} & Space group & Supercell & Full formula of unsubstituted structure 
    & Wyckoff(s){\footnotemark[2]} & \# Substitutions{\footnotemark[3]} & \# Equivalent{\footnotemark[4]} 
    & $T_{\supercell}${\footnotemark[5]} & $T_{\shry}${\footnotemark[5]}
    \\    \hline
    \endfirsthead
    \multicolumn{10}{c}%
    {\tablename\ \thetable\ -- \textit{Continued from previous page}} \\
    \hline
    Reduced formula & Code{\footnotemark[1]} & Space group & Supercell & Full formula of unsubstituted structure 
    & Wyckoff(s){\footnotemark[2]} & \# Substitutions{\footnotemark[3]} & \# Equivalent{\footnotemark[4]}
    & $T_{\supercell}${\footnotemark[5]} & $T_{\shry}${\footnotemark[5]}
    \\    \hline
    \endhead
    \hline \multicolumn{10}{c}{\textit{Continued on next page}} \\
    \endfoot
    \hline
    \endlastfoot
H$_{{26}}$C$_{{26}}$SN$_{{2}}$O$_{{3}}$ & 7217822 & $P1$ (1) & 3$\times$2$\times$1 & H$_{156}$C$_{156}$S$_{6}$N$_{12}$O$_{18}$ & H$_{6}$(1a) $\rightarrow$ Fr$_{1}$Cs$_{2}$Rb$_{1}$Na$_{2}$, H$_{6}$(1a) $\rightarrow$ Cs$_{1}$K$_{1}$Na$_{3}$H$_{1}$, C$_{6}$(1a) $\rightarrow$ Ge$_{3}$Fl$_{2}$Si$_{1}$ & 1,296,000 & 216,000 & 2.560 & 114.890 \\
CaV$_{{2}}$P$_{{2}}$(H$_{{4}}$O$_{{7}}$)$_{{2}}$ & 9012519 & $P1$ (1) & 2$\times$2$\times$2 & Ca$_{8}$V$_{16}$P$_{16}$H$_{64}$O$_{112}$ & O$_{8}$(1a) $\rightarrow$ Lv$_{4}$Te$_{1}$Se$_{2}$S$_{1}$, O$_{8}$(1a) $\rightarrow$ Po$_{1}$Te$_{1}$S$_{5}$O$_{1}$ & 282,240 & 35,280 & 0.635 & 15.655 \\
CoH$_{{18}}$C$_{{14}}$N$_{{2}}$O$_{{5}}$ & 8000383 & $P1$ (1) & 3$\times$2$\times$2 & Co$_{12}$H$_{216}$C$_{168}$N$_{24}$O$_{60}$ & H$_{12}$(1a) $\rightarrow$ K$_{2}$Li$_{10}$, H$_{12}$(1a) $\rightarrow$ Fr$_{3}$Rb$_{2}$Na$_{3}$Li$_{4}$ & 18,295,200 & 1,524,600 & 26.585 & 14.120 \\
H$_{{11}}$C$_{{14}}$NO$_{{4}}$ & 7151465 & $P\bar1$ (2) & 2$\times$2$\times$1 & H$_{176}$C$_{224}$N$_{16}$O$_{64}$ & C$_{8}$(2i) $\rightarrow$ Fl$_{1}$Sn$_{2}$Ge$_{5}$, C$_{8}$(2i) $\rightarrow$ Fl$_{3}$Si$_{2}$Ge$_{3}$, H$_{8}$(2i) $\rightarrow$ Fr$_{5}$Cs$_{3}$ & 5,268,480 & 658,560 & 9.570 & 33.355 \\
P$_{{4}}$H$_{{15}}$(RuO$_{{10}}$)$_{{2}}$ & 7049930 & $P\bar1$ (2) & 2$\times$3$\times$1 & P$_{24}$H$_{90}$Ru$_{12}$O$_{120}$ & H$_{12}$(2i) $\rightarrow$ Cs$_{3}$H$_{9}$, O$_{12}$(2i) $\rightarrow$ Lv$_{5}$Se$_{7}$, H$_{12}$(2i) $\rightarrow$ Cs$_{9}$K$_{3}$ & 38,332,800 & 3,194,400 & 25.210 & 18.720 \\
H$_{{17}}$C$_{{20}}$SBr(NO)$_{{4}}$ & 7154816 & $P\bar1$ (2) & 1$\times$3$\times$1 & H$_{102}$C$_{120}$S$_{6}$Br$_{6}$N$_{24}$O$_{24}$ & C$_{6}$(2i) $\rightarrow$ Si$_{2}$Pb$_{1}$C$_{3}$, C$_{6}$(2i) $\rightarrow$ Si$_{2}$Sn$_{2}$Ge$_{1}$Pb$_{1}$ & 10,800 & 1,800 & 0.415 & 2.530 \\
\end{longtable}
  \footnotetext[1]{Unique ID in Crystallography Open Database (COD)~{\cite{2009GRA}}.}
  \footnotetext[2]{Substituted Wyckoff position(s).}
  \footnotetext[3]{The total amount of substituted pattern including symmetry-equivalent strutures.}
  \footnotetext[4]{The number of symmetry-inequivalent structures.}
  \footnotetext[5]{The CPU times (sec.) of \shry\ and \supercell\ were measured on 
  Intel{\textsuperscript{\textregistered}} Xeon{\textsuperscript{\textregistered}} G-6242 (2.8~GHz) processor, 
  where time-consuming I/O operations were disabled. The distributed binary
  \supercell\ program was used for the benchmark test.
  The CPU times were measured by the {\textsc{time}} command twice and averaged.}
\end{landscape}
\end{center}
\endgroup

\end{document}